\documentclass{PoS}

\newcommand{\beq}{\begin{equation}}
\newcommand{\eeq}{\end{equation}}
\newcommand{\beqa}{\begin{eqnarray}}
\newcommand{\eeqa}{\end{eqnarray}}
\newcommand{\be}{\begin{equation}}
\newcommand{\ee}{\end{equation}}

\newcommand{\vp}{\vec{p}}
\newcommand{\vq}{\vec{q}}
\newcommand{\vk}{\vec{k}}
\newcommand{\op}{\omega_{p}}
\newcommand{\oq}{\omega_{q}}
\newcommand{\ok}{\omega_{k}}

\newcommand{\intk}{\sum_{n_k} \int \frac{d^3k}{(2\pi)^3}}

\def\Eq#1{Eq.~(\ref{#1})}

\title{Quark condensates and the deconfinement transition}

\ShortTitle{Quark condensates and the deconfinement transition}

\author{\speaker{Christian S. Fischer}\thanks{This work has 
been supported by the Helmholtz Young Investigator 
Grant VH-NG-332 and by the Helmholtz Alliance HA216-TUD/EMMI.}\\
         \\
 Institute for Nuclear Physics, 
 Darmstadt University of Technology,\\ 
 Schlossgartenstra{\ss}e 9, 64289 Darmstadt, Germany\\
 GSI Helmholtzzentrum f\"ur Schwerionenforschung GmbH,\\ 
 Planckstra{\ss}e 1,  64291 Darmstadt, Germany.\\
        E-mail: \email{christian.fischer@physik.tu-darmstadt.de}}

\author{Jens~A.~Mueller\\
 Institute for Nuclear Physics, 
 Darmstadt University of Technology,\\ 
 Schlossgartenstra{\ss}e 9, 64289 Darmstadt, Germany\\
        E-mail: \email{jens.mueller@physik.tu-darmstadt.de}}

\abstract{In this talk we present results on the chiral and the 
deconfinement transition of quenched QCD from Dyson-Schwinger 
equations. We determine the ordinary quark condensate 
signaling the chiral transition and the dual quark condensate 
signaling the deconfinement transition from the Landau gauge
quark propagator evaluated at generalized boundary conditions. 
We find only slightly different transition temperatures at 
finite quark masses, whereas the transition temperatures 
coincide in the chiral limit.}

\FullConference{5th International Workshop on Critical Point and Onset of Deconfinement - CPOD 2009,\\
		 June 08 - 12 2009\\
		 Brookhaven National Laboratory, Long Island, New York, USA}

\begin{document}

\section{Introduction}
The QCD phase diagram is a matter of intense investigation
from both, theory and experiment. Some of the pressing open 
questions discussed at this conference are the presence or
absence of a critical point \cite{deForcrand:2006pv}, the 
possibility of a confined chirally symmetric ('quarkyonic') 
phase \cite{McLerran:2007qj} and the (non-)coincidence of the
chiral and the deconfinement transition at zero chemical
potential \cite{Bazavov:2009zn,Aoki:2009sc}. Answers to these
questions certainly require nonperturbative approaches to QCD.
On the other hand, the considerable complexity of these questions 
suggests that one approach alone is hardly capable to provide
all answers. Instead it seems promising to combine the various 
available methods in order to balance their respective strengths 
and weaknesses. 

Lattice Monte-Carlo simulations are well behaved
at zero or imaginary chemical potential, but encounter the 
notorious sign problem when it comes to real chemical potential.
One approach to overcome this problem, extrapolation from zero 
chemical potential by Taylor-expansion methods 
(see e.g. \cite{Ejiri:2003dc}), has been questioned recently on 
the basis of the failure of corresponding extrapolations in model 
calculations \cite{taylor}. These models, notably the 
Polyakov-NJL model and the Polyakov-quark-meson model 
(see e.g. \cite{Schaefer:2007pw,Rossner:2007ik,Fukushima:2008is}
and refs. therein) have been employed frequently to explore
the details of the QCD phase diagram at zero and finite chemical 
potential. Their success is demonstrated e.g. by the quantitative 
reproduction of the lattice equation of state at zero chemical 
potential. Furthermore they serve as a formidable qualitative 
playground to explore scenarios for the details of the interplay 
between the chiral and the deconfinement transition. Nevertheless, 
it is hard to see how the various model parameters can be 
constrained enough to arrive at quantitative predictions as e.g. 
the location of a possible critical point \cite{Stephanov:2007fk}. 

A third class of approaches are functional methods involving the 
renormalization group equations \cite{Berges:2000ew} and/or 
Dyson-Schwinger equations \cite{Roberts:2000aa,Fischer:2006ub} of 
QCD. In the past years much progress has been made in extending 
these methods to finite temperature and/or chemical potential,
see e.g. 
\cite{Bender:1996bm,Roberts:2000aa,Braun:2006jd,Braun:2007bx,%
Braun:2008pi,Fischer:2009wc,fimu,heid}.
Most of the earlier works involving functional methods 
concentrated on the chiral aspects of the QCD transition. 
Recently, methods became available that also take into account the 
deconfining aspect \cite{Bender:1996bm,Braun:2007bx,Fischer:2009wc,fimu,heid}. 
In this talk we report on results from one particular method that 
extracts the deconfinement transition temperature from the 
properties of the quark propagator at generalized boundary 
conditions. This method has been introduced originally within the 
lattice framework \cite{Gattringer:2006ci,Bilgici:2008qy}
and adapted to functional methods in \cite{Fischer:2009wc,fimu,heid}. 
The quantity signaling the deconfinement transition is the dual 
quark condensate (or 'dressed Polyakov loop'). It transforms under 
center transformations exactly like the ordinary Polyakov loop and 
is therefore an order parameter in the limit of infinitely heavy quarks. 
This quantity is furthermore interesting from a formal perspective 
since it relates both the chiral and the deconfinement transition 
to the spectral properties of the Dirac operator 
\cite{Gattringer:2006ci,Synatschke:2008yt}. 

In the following we first recall the defining equations for the 
ordinary and the dual quark condensate, then summarize the truncation
scheme used in our DSE calculations before we discuss our results
for the chiral and deconfinement phase transition.

\section{The dual quark condensate}

Consider the quark propagator $S(\vp,\op)$ at finite temperature 
given by the tensor decomposition
\beq \label{quark}
S(\vp,\op) = \left[i \gamma_4\, \op C(\vp,\op) 
+ i \gamma_i \, p_i A(\vp,\op) + B(\vp,\op)\right]^{-1} \,,
\eeq
with vector and scalar quark dressing functions $C,A,B$. At 
physical, antiperiodic boundary conditions the corresponding 
Matsubara frequencies are given by $\op(n_t) = (2\pi T)(n_t + 1/2)$. 
The ordinary quark condensate can be extracted from the trace of 
the quark propagator by
\beq  \label{cond}
 \langle\bar{\psi}\psi \rangle_{\varphi} = 
 Z_m\, Z_2\, N_c\, T\sum_{n_t}\int\frac{d^3p}{(2\pi)^3}\,
 \textrm{tr}_D\,S(\vp,\op)\,.
\eeq
For vanishing bare quark masses this integral is well-behaved
and delivers the chiral condensate, whereas at finite bare quark 
masses it is quadratically divergent and needs to be properly 
regularized.

Consider now non-standard, $U(1)$-valued boundary conditions 
in the temporal direction established by the equation
$\psi(\vec{x},1/T) = e^{i \varphi} \psi(\vec{x},0)$ for the
quark field $\psi$ with the boundary angle $\varphi \in [0,2\pi[$.
For the physical antiperiodic fermion boundary conditions we 
have $\varphi=\pi$, whereas $\varphi=0$ corresponds to periodic 
boundary conditions. The corresponding Matsubara frequencies
are given by $\op(n_t,\varphi) = (2\pi T)(n_t+\varphi/2\pi)$.
In this non-standard framework one can evaluate a quark
condensate by 
\beq \label{trace}
 \langle\bar{\psi}\psi \rangle_{\varphi} = 
 Z_2\, N_c\, T\sum_{n_t}\int\frac{d^3p}{(2\pi)^3}\,
 \textrm{tr}_D\,S(\vec{p},\omega_p(\varphi))
\eeq
with the conventional quark condensate obtained for $\varphi=\pi$
and multiplication with $Z_m$. It has been shown in 
\cite{Bilgici:2008qy} that the Fourier-transform 
\beq \label{dual}
\Sigma_1 = \int_0^{2\pi} \, \frac{d \varphi}{2\pi} \, e^{-i\varphi}\, 
\langle \overline{\psi} \psi \rangle_\varphi
\eeq
of this $\varphi$-dependent condensate delivers a quantity that
transforms under center transformations exactly like the Polyakov-loop
and is therefore an order parameter for the deconfinement transition.
This quantity $\Sigma_1$ is called the dual condensate or dressed 
Polyakov loop.

The relation of $\Sigma_1$ to the ordinary Polyakov-loop becomes
transparent in the following loop expansion of the $\varphi$-dependent
condensate:
\beq \label{loop}
\langle \overline{\psi} \psi \rangle_\varphi =  
\sum_{l \in \mathcal{L}} \frac{e^{i\varphi n(l)}}{(a m)^{|l|}} U(l) \,.
\eeq
Here $\mathcal{L}$ denotes the set of all closed loops $l$ with
length $|l|$ on a lattice with lattice spacing $a$. Furthermore $m$ 
is the quark mass.  $U(l)$ stands for the chain of link variables in 
a loop $l$ multiplied with appropriate sign and normalization factors, 
see Ref.~\cite{Bilgici:2008qy} for details. Each loop that closes 
around the temporal boundary picks up factors of $e^{\pm i\varphi}$ 
according to its winding number $n(l)$. Correspondingly, the 
Fourier-transform in \Eq{dual} projects out exactly those loops which
wind once around the temporal direction of the lattice (therefore the 
notation $\Sigma_1$). In the limit of heavy quark masses long loops
are suppressed by $1/m^{|l|}$ and therefore only the straight line
along the temporal direction of the lattice survives; the dual
condensate is then equal to the ordinary Polyakov-loop.

An interesting property of the dual condensate $\Sigma_1$ is the fact
that it can be evaluated with functional methods \cite{Fischer:2009wc},
as we will see in the following.  

\section{The Dyson-Schwinger equation for the quark propagator at finite
temperature}

%%%%%%%%%%%%%%%%%%%%%%%%%%%%%%%%%%%%%%%%%%%%%%%%%%%%%%%%%%%%%%%%%%%%%%%%
\begin{figure}[t]
\centerline{\includegraphics[width=0.7\columnwidth]{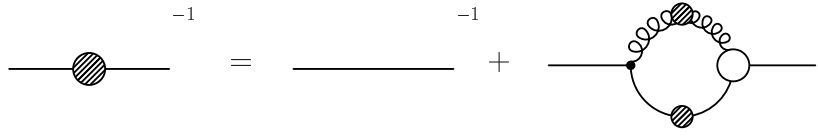}}
\caption{The Dyson-Schwinger equation for the quark propagator. Filled circles
denote dressed propagators whereas the empty circle stands for the dressed
quark-gluon vertex.
\label{fig:quarkDSE}}
\end{figure}
%%%%%%%%%%%%%%%%%%%%%%%%%%%%%%%%%%%%%%%%%%%%%%%%%%%%%%%%%%%%%%%%%%%%%%%%%
The Dyson-Schwinger equation for the quark propagator \Eq{quark} is
displayed diagrammatically in Fig.~\ref{fig:quarkDSE}. At
finite temperature $T$ it is given by 
\beqa \label{DSE}
S^{-1}(p) = Z_2 \, S^{-1}_0(p) 
-  C_F\, Z_{1f}\, g^2 T \intk \, \gamma_{\mu}\, S(k) \,\Gamma_\nu(k,p) \,
D_{\mu \nu}(p-k) \,, \label{quark_t}
\eeqa
with $p=(\vp,\op)$ and $k=(\vk,\ok)$ and renormalization factors
$Z_2$ and $Z_1f$. Here $D_{\mu \nu}$ denotes the (transverse) gluon 
propagator in Landau gauge and $\Gamma_\nu$ the quark-gluon vertex. 
The bare quark propagator is given by $S^{-1}_0(p) = i \gamma \cdot p + m$. 
The Casimir factor $C_F = (N_c^2-1)/N_c$ stems from the color trace; here 
we only consider the gauge group $SU(2)$. The quark dressing functions 
$A,B,C$ can be extracted from Eq.~(\ref{DSE}) 
by suitable projections in Dirac-space. 

In order to solve this equation we have to specify explicit expressions
for the gluon propagator and the quark-gluon vertex. At finite temperatures
the tensor structure of the gluon propagator contains two parts, one 
transversal and one longitudinal to the heat bath. The propagator is then 
given by ($q=(\vq,\oq)$)
\begin{eqnarray}
D_{\mu\nu}(q) = \frac{Z_T(q)}{q^2} P_{\mu \nu}^T(q) 
                    +\frac{Z_L(q)}{q^2} P_{\mu \nu}^L(q) 
\end{eqnarray} 
with transverse and longitudinal projectors 
\beqa
P_{\mu\nu}^T(q) = 
   \left(\delta_{i j}-\frac{q_i q_j}{\vq^2}\right) 
   \delta_{i\mu}\delta_{j\nu}\,, \hspace*{2cm}
P_{\mu\nu}^L(q) = P_{\mu \nu}(q) - P_{\mu \nu}^T(q) \,,
\eeqa 
with ($i,j=1 \dots 3$). The transverse dressing $Z_T(\vq,\oq)$ is also known 
as magnetic dressing function of the gluon, whereas the longitudinal component 
$Z_L(q)$ is called electric dressing function of the gluon propagator. At
zero temperatures Euclidean $O(4)$-invariance requires both dressing functions 
to agree, i.e. $Z_T(q)=Z_L(q)=Z(q)$.

\begin{figure}[t]
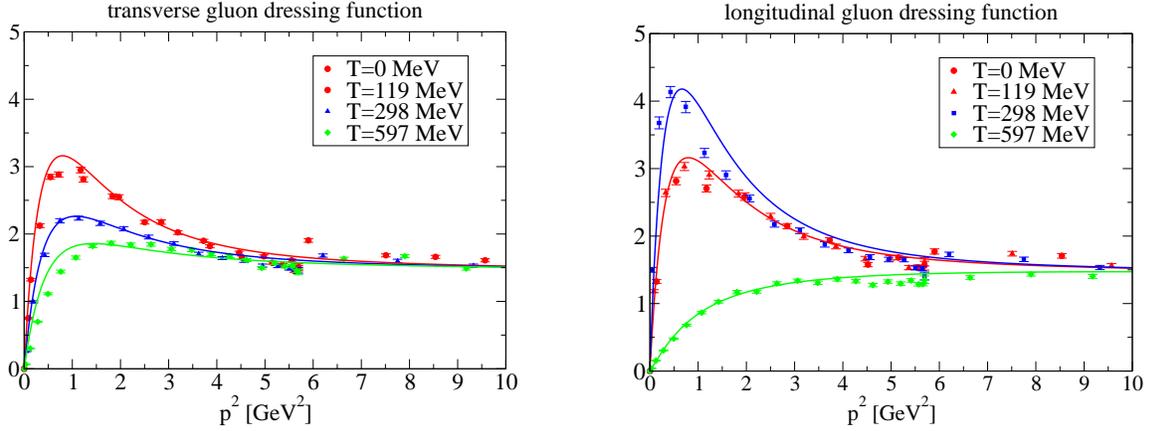

\includegraphics[width=0.45\columnwidth]{glue-trans.eps}\hfill
\includegraphics[width=0.45\columnwidth]{glue-long.eps}
\caption{Quenched $SU(2)$ lattice results \cite{Cucchieri:2007ta} for the 
transverse dressing function $Z_T(q)$ and the longitudinal dressing
function $Z_L(q)$ of the gluon propagator together with the fit 
functions \cite{Fischer:2009wc}.}
\label{fig:lattglue}
\end{figure}

The temperature dependence of the gluon propagator can be inferred from
recent lattice calculations. The results of Ref.~\cite{Cucchieri:2007ta} 
are shown in Fig.~\ref{fig:lattglue}. The temperature effects on both the 
magnetic and electric dressing functions are such that there are almost no 
effects when comparing the $T=0$ result with $T=119$ MeV. Further increasing 
the temperature to $T=298$ MeV and $T=597$ MeV significantly decreases the 
bump in the magnetic dressing function around $p^2=1$ GeV$^2$. There is no 
indication that this decrease takes special notice of the critical temperature
$T_c \approx 300$ MeV for quenched QCD with gauge group $SU(2)$. The
opposite seems to be true for the electric part of the propagator. Here from 
$T=119$ MeV to $T=300$ MeV one observes a clear increase of the bump in the 
dressing function $Z_L(q)$ and a subsequent decrease when the temperature
is further raised to $T=597$ MeV. Pending further investigation it seems 
reasonable to assume that the maximum of the bump is reached at or around the 
critical temperature $T_c \approx 300$ MeV.

%%%%%%%%%%%%%%%%%%%%%%%%%%%%%%%%%%%%%%%%%%%%%%%%%%%%%%%%%%%%%%%%%%%%%%%%
%%%%%%%%%%%%%%%%%%%%%%%%%%%%%%%%%%%%%%%%%%%%%%%%%%%%%%%%%%%%%%%%%%%%%%%%%

Although the lattice data still have considerable systematic errors 
\cite{Cucchieri:2007ta} they may very well correctly represent the qualitative 
temperature dependence of the gluon propagator. We therefore use a 
temperature dependent (qualitative) fit to the data as input into the DSE; 
this fit is also displayed in Fig.~\ref{fig:lattglue} (straight lines). 
The fit functions are described in detail in Refs.~\cite{Fischer:2009wc,fimu}
and shall not be repeated here for brevity. Note, however, that we also 
inherit the scale determined on the lattice using the string tension 
$\sqrt{\sigma}=0.44$ GeV \cite{Cucchieri:2007ta}. 

For the quark-gluon vertex with gluon momentum $q=(\vq,\oq)$ and the quark 
momenta $p=(\vp,\op),k=(\vk,\ok)$ we employ the following temperature 
dependent model
\beqa \label{vertexfit}
\Gamma_\nu(q,k,p) &=& \widetilde{Z}_{3}\left(\delta_{4 \nu} \gamma_4 
\frac{C(k)+C(p)}{2}
+  \delta_{j \nu} \gamma_j 
\frac{A(k)+A(p)}{2}
\right)\times \nonumber\\
&&\left( 							
\frac{d_1}{d_2+q^2} 			
 + \frac{q^2}{\Lambda^2+q^2}
\left(\frac{\beta_0 \alpha(\mu)\ln[q^2/\Lambda^2+1]}{4\pi}\right)^{2\delta}\right) \,,
\eeqa 
where $\delta=-9/44$ is the anomalous dimension of the vertex. 
The dependence of the vertex on the quark dressing
functions $A$ and $C$ is motivated by the Slavnov-Taylor identity for the 
vertex. The remaining fit function is purely phenomenological, see 
e.g. \cite{Fischer:2008wy} where an elaborate version of such an ansatz 
has been used to describe meson observables. The parameters are given by 
$d_1 = 7.6 \,\mbox{GeV}^2$ and $d_2=0.5 \,\mbox{GeV}^2$. A moderate variation 
of these parameters shifts the critical temperatures of both, 
the chiral and the deconfinement transition but leaves all qualitative 
aspects of the results presented below unchanged. 

The truncation scheme described above has the merit to explicitly implement
a realistic temperature dependence of the gluon propagator and the 
quark-gluon vertex beyond simple ansaetze, see e.g. 
\cite{Bender:1996bm,Roberts:2000aa,Maris:2000ig,Horvatic:2007qs} 
for previous approaches. The explicit expressions of the resulting DSEs 
for the quark dressing functions together with the details of our
numerical method are given in Ref.~\cite{fimu}. 
 
\section{Numerical results}
%%%%%%%%%%%%%%%%%%%%%%%%%%%%%%%%%%%%%%%%%%%%%%%%%%%%%%%%%%%%%%%%%%%%%%%%
\begin{figure}[t]
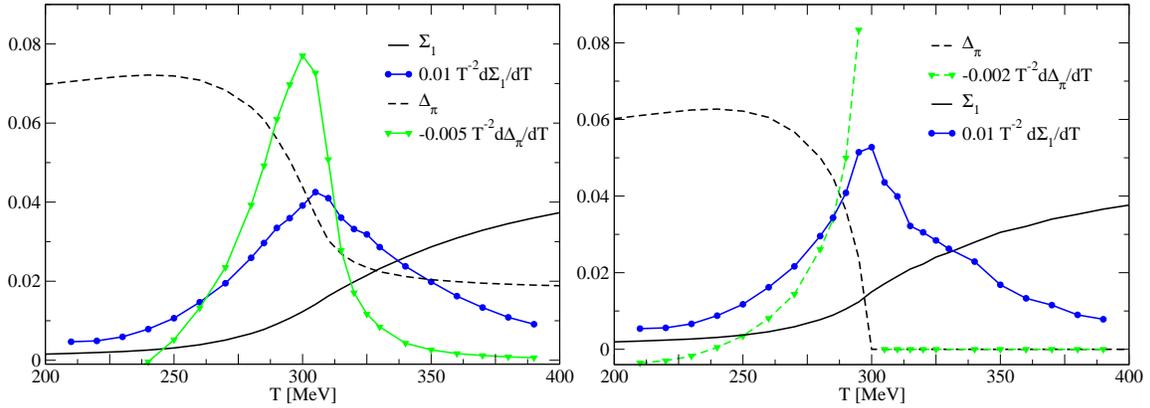

\includegraphics[width=0.5\columnwidth]{cond_dual_cond_vs_T_m10MeV.eps}\hfill
\includegraphics[width=0.5\columnwidth]{Kond_Dual-Kond_vs_Temp_chiral.eps}
\caption{
Left diagram: Temperature dependence of the dressed Polyakov-loop 
$\Sigma_1$ and the conventional quark condensate 
$\Delta_\pi \equiv \langle \overline{\psi} \psi \rangle_{\varphi=\pi}$  
together with their derivatives for $m = 10 \,\mbox{MeV}$.
Right diagram: The same quantities in the chiral limit.}
\label{res1}
\end{figure}
%%%%%%%%%%%%%%%%%%%%%%%%%%%%%%%%%%%%%%%%%%%%%%%%%%%%%%%%%%%%%%%%%%%%%%%%%

In Fig.~\ref{res1} we display our numerical results for the
ordinary and the dual quark condensate together with their (normalized) 
temperature derivatives once evaluated for a quark mass of 
$m = 10 \,\mbox{MeV}$ and once evaluated in the chiral limit. One clearly
sees the difference in the chiral transition: whereas at finite bare quark
mass we encounter a crossover the transition changes into a second order
phase transition in the chiral limit. In the first case the corresponding 
temperature derivative shows a peak at $T_c = 301(2)$ MeV, whereas it
diverges at $T_c = 298(1)$ MeV in the second case. We also extracted the
corresponding transition temperatures from the chiral susceptibility
\beq
% \chi=\partial/(\partial\, m)~\langle\bar{\psi} \psi\rangle\,, \hspace*{2cm}
 \chi_R=m^2\frac{\partial}{\partial m}
 \Big(\langle\bar{\psi}\psi\rangle_T-\langle\bar{\psi}\psi\rangle_{T=0}\Big)\,. 
 \label{def2}
\eeq
The results for quark mass $m = 10 \,\mbox{MeV}$ are given in table \ref{crit_temp}.
%%%%%%%%%%%%%%%%%%%%%%%%%%%%%%%%%%%%%%%%%%%%%%%%%%%%%%%%%%%%%%%%%%%%%%%%%%%%%%
\begin{table}[b]
\begin{center}
\begin{tabular}{c|c|c|c}
 $T_c$  & $T_{\chi_R/T^4}$ & $T_{\chi_R}$ & $T_{dec}$   \\\hline\hline
$301(2)$ & $304(1)$         & $305(1)$     & $308(2)$
\end{tabular}
\caption{Transition temperatures for the chiral and deconfinement transition
for quark mass $m = 10 \,\mbox{MeV}$.} \label{crit_temp}
\end{center}
\end{table}
%%%%%%%%%%%%%%%%%%%%%%%%%%%%%%%%%%%%%%%%%%%%%%%%%%%%%%%%%%%%%%%%%%%%%%%%%%%%%%

The corresponding transition temperature for the deconfinement transition
can be read off the dual quark condensate (or dressed Polyakov loop). At finite
quark mass and in the chiral limit we observe a distinct rise in the dual 
condensate around $T \approx 300$ MeV. The corresponding (normalized)
temperature derivative shows peaks at $T_{dec} = 308(2)$ MeV for 
quark mass $m = 10 \,\mbox{MeV}$. In the chiral limit this peak moves
to $T_{dec} =299(3)$ MeV. 

In general we note that the chiral and 
deconfinement transition are close together. There are a few MeV difference 
between the different transition temperatures for the crossover at finite 
quark masses, whereas both transitions occur at the same temperature (within errors)
in the chiral limit. These findings agree with early expectations from lattice
simulations \cite{Karsch:1998ua}.

Furthermore we wish to emphasize that the present calculation, although carried
out with quenched lattice results for the gluon propagator, is in itself
not strictly quenched: our ansatz for the quark-gluon vertex is too simple
to strictly represent the quenched theory. This can be seen from the fact that
the dressed Polyakov-loop is not strictly zero below the deconfinement
transition. Consequently we do not observe the second order deconfinement
phase transition expected from quenched $SU(2)$ Yang-Mills theory. Note,
however, that even if our vertex were strictly quenched it is not clear 
whether the lattice input for the gluon propagator is precise enough to 
allow for an observation of the second order phase transition.

\begin{figure}[t]
\begin{center}
\includegraphics[width=0.5\columnwidth]{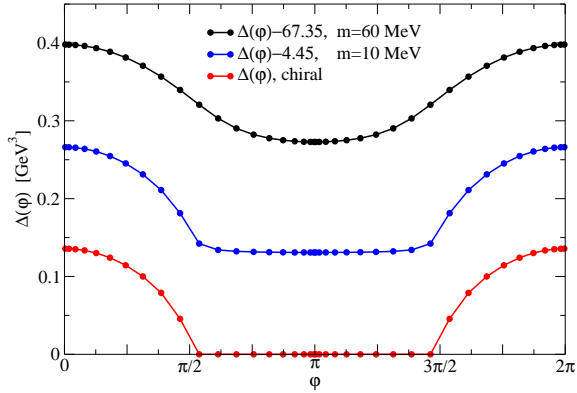}
\end{center}
\caption{
Angular dependence of the quark condensate evaluated at two different 
quark masses and in the chiral limit at $T=400$ MeV.
}
\label{res2}
\end{figure}

The details of the mass dependence of the $\varphi$-dependent condensate 
are studied in Fig.~\ref{res2}. We compare the angular dependence of the 
condensate at $T=400$ MeV for two different quark masses and in the chiral 
limit. We clearly see a broadening in the central dip of the graphs with 
decreasing quark mass. This can be readily understood from the loop expansion 
of the quark condensate, Eq.(\ref{loop}). 
At sufficiently large quark masses large loops are 
suppressed by powers of $1/m$. As a result only loops winding once around 
the torus should contribute in \Eq{loop} and the resulting angular behavior 
of the condensate should be proportional to $\cos(\varphi)$. Indeed, this 
is what we see: the result for our largest quark mass can be well fitted 
by only few terms in an expansion 
$\Delta(\varphi) = \sum_{n=0}^N a_n \cos(n\varphi)$ and the first term
is by far the largest contribution. For smaller quark masses we observe
also sizeable contributions from terms $\cos(n\varphi)$ with $n>1$. In the 
plot, these contributions are responsible for the flat area around the 
antiperiodic boundary angle $\varphi=\pi$. Approaching the chiral limit 
this area becomes flatter and finally develops a derivative discontinuity
at two finite values of $\varphi = \pi \pm L$. These indicate the breakdown 
of the loop expansion \Eq{loop} in the chiral limit. 

%%%%%%%%%%%%%%%%%%%%%%%%%%%%%%%%%%%%%%%%%%%%%%%%%%%%%%%%%%%%%%%%%%%%%%%%%%%%%
%%%%%%%%%%%%%%%%%%%%%%%%%%%%%%%%%%%%%%%%%%%%%%%%%%%%%%%%%%%%%%%%%%%%%%%%%%%%%

\begin{figure}[t]
\includegraphics[width=\columnwidth]{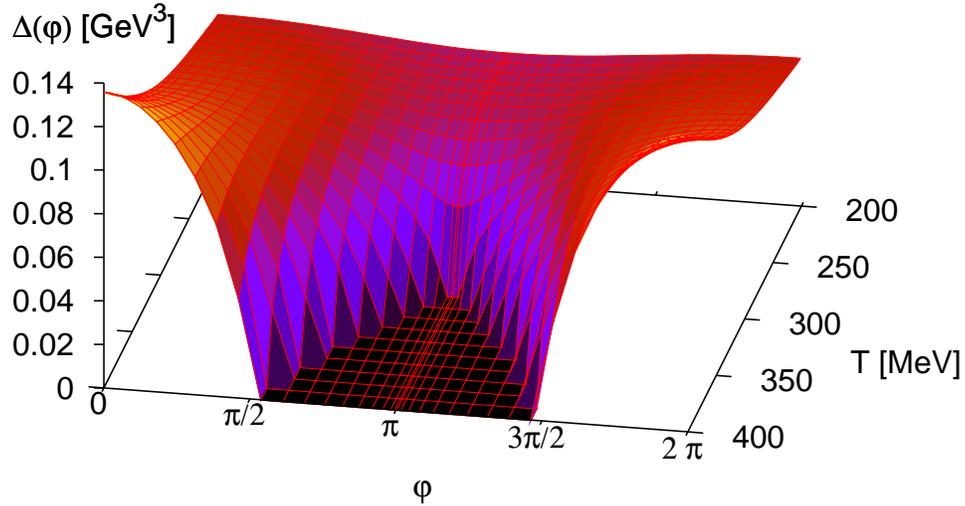}
\caption{
A 3d-plot of the angular and temperature dependence of the chiral 
quark condensate.}
\label{res3}
\end{figure}

Finally we show the angular and temperature dependence of the $\varphi$-dependent 
condensate $\Delta_{\varphi}(T)$ in Fig.~\ref{res3}. The 3d-plot clearly shows
the different evolution of the condensate at varying boundary conditions.
Whereas at physical antiperiodic boundary angle $\varphi=\pi$ we observe the
second order chiral phase transition also shown in Fig.~\ref{res1}, we find a 
monotonically rising condensate at the periodic boundary conditions 
$\varphi=0$. For larger temperatures (not shown in the plot) we can extract a 
quadratic rise of the $\varphi$-dependent condensate $\Delta_{\varphi=0}(T)$, 
\beq
\Delta_{\varphi=0}(T) \sim T^2  \hspace{1cm}\mbox{for} \hspace{2mm}T\gg T_c \,.
\eeq
This behavior can also be extracted analytically from Eqs.~(\ref{DSE}) and (\ref{cond})
for the quark propagator and the quark condensate as shown in the appendix of 
Ref.~\cite{fimu}. Around the physical value of $\varphi=\pi$ we see a
plateau with $\Delta_{\varphi}(T) = 0$ that gets broader with increasing
temperature. The width of this plateau seems to settle at a finite value
smaller than $2\pi$ for $T>2T_c$; however from the available results we can
neither show nor exclude that it approaches $2\pi$ very slowly for 
$T \rightarrow \infty$.

%%%%%%%%%%%%%%%%%%%%%%%%%%%%%%%%%%%%%%%%%%%%%%%%%%%%%%%%%%%%%%%%%%%%%%%%%%%%%
%%%%%%%%%%%%%%%%%%%%%%%%%%%%%%%%%%%%%%%%%%%%%%%%%%%%%%%%%%%%%%%%%%%%%%%%%%%%%

\section{Summary}

In this talk we addressed the chiral and the deconfinement transition 
of quenched QCD. We showed results for the order parameter for the 
chiral transition, the quark condensate, and an order parameter for 
the deconfinement transition, the dressed Polyakov loop extracted
from the Landau gauge quark propagator evaluated at a continuous
range of boundary conditions for the quark fields. We found almost
coinciding transition temperatures for the chiral and the deconfinement
transition at a moderate quark mass of the order of an up-quark.
In the chiral limit the two transitions coincide within error.
We find a second order chiral phase transition at $T_{\chi_R/T^4} = 298(1)$ 
MeV and a similar temperature for the deconfinement transition, 
$T_{dec} =299(3)$ MeV. It is worth to emphasize again that both
transition temperatures are extracted from the properties of the
quark propagator, respectively the underlying properties of the
Dirac operator.

The framework used in this work is quenched $SU(2)$ Yang-Mills theory.
Our transition temperature may be translated into the corresponding ones
of quenched $SU(3)$ QCD using the relations $T_c/\sqrt{\sigma}=0.709$
($SU(2)$) and $T_c/\sqrt{\sigma}=0.646$ ($SU(3)$) between the respective 
critical temperatures and the string tension \cite{Fingberg:1992ju}. 
The resulting transition temperature is then 
$T_{\chi_R/T^4} \approx T_{dec} \approx 272$ MeV in the chiral limit.
In order to work in the full, unquenched theory we would have to take
into account quark-loop effects in the gluon propagator and meson effects
in the quark-gluon vertex \cite{Fischer:2008wy}. These effects will shift
the transition temperatures below $T=200$ MeV, see \cite{Bazavov:2009zn,Aoki:2009sc}
for latest results for $N_f=2+1$ quark flavors. As concerns the dual 
condensate and scalar dressing function in the unquenched formulation 
one needs to carefully take into account effects due to the 
Roberge-Weiss symmetry \cite{Roberge:1986mm}. This is because of the formal 
similarity of the continuous boundary conditions for the quark field to an 
imaginary chemical potential, see \cite{heid} for details. 

Note: After completion of this proceedings contribution also first results
for the dual condensate evaluated in the PNJL-model became available 
\cite{Kashiwa:2009ki}.

{\bf Acknowledgments}\\
We thank Falk Bruckmann, Christof Gattringer, Erwin Laermann,
Jan Pawlowski, Rob Pisarski, Lorenz von Smekal and Wolfgang Soeldner 
for discussions. We are grateful to Axel Maas 
for discussions and for making the lattice data of 
Ref.~\cite{Cucchieri:2007ta} available. This work has 
been supported by the Helmholtz Young Investigator 
Grant VH-NG-332 and by the Helmholtz Alliance HA216-TUD/EMMI.


\begin{thebibliography}{99}

%\cite{deForcrand:2006pv}
\bibitem{deForcrand:2006pv}
  P.~de Forcrand and O.~Philipsen,
  %``The chiral critical line of N_f=2+1 QCD at zero and non-zero baryon
  %density,''
  JHEP {\bf 0701} (2007) 077
  [arXiv:hep-lat/0607017];
  %%CITATION = JHEPA,0701,077;%%
%\cite{Philipsen:2009yg}
%\bibitem{Philipsen:2009yg}
  O.~Philipsen,
  %``Towards the chiral critical surface of QCD,''
  arXiv:0907.4668 [hep-ph].
  %%CITATION = ARXIV:0907.4668;%%

%\cite{McLerran:2007qj}
\bibitem{McLerran:2007qj}
  L.~McLerran and R.~D.~Pisarski,
  %``Phases of Cold, Dense Quarks at Large N_c,''
  Nucl.\ Phys.\  A {\bf 796} (2007) 83
  [arXiv:0706.2191 [hep-ph]];
  %%CITATION = NUPHA,A796,83;%%
%\cite{McLerran:2008ua}
%\bibitem{McLerran:2008ua}
  L.~McLerran, K.~Redlich and C.~Sasaki,
  %``Quarkyonic Matter and Chiral Symmetry Breaking,''
  Nucl.\ Phys.\  A {\bf 824}, 86 (2009)
  [arXiv:0812.3585 [hep-ph]].
  %%CITATION = NUPHA,A824,86;%%

%\cite{Bazavov:2009zn}
\bibitem{Bazavov:2009zn}
  A.~Bazavov {\it et al.},
  %``Equation of state and QCD transition at finite temperature,''
  arXiv:0903.4379 [hep-lat].
  %%CITATION = ARXIV:0903.4379;%%

%\cite{Aoki:2009sc}
\bibitem{Aoki:2009sc}
  Y.~Aoki {\it et al.}, 
  %S.~Borsanyi, S.~Durr, Z.~Fodor, S.~D.~Katz, S.~Krieg and K.~K.~Szabo,
  %``The QCD transition temperature: results with physical masses in the
  %continuum limit II,''
  arXiv:0903.4155 [hep-lat].
  %%CITATION = ARXIV:0903.4155;%%

%\cite{Ejiri:2003dc}
\bibitem{Ejiri:2003dc}
  S.~Ejiri, C.~R.~Allton, S.~J.~Hands, O.~Kaczmarek, F.~Karsch, E.~Laermann and C.~Schmidt,
  %``Study of QCD thermodynamics at finite density by Taylor expansion,''
  Prog.\ Theor.\ Phys.\ Suppl.\  {\bf 153} (2004) 118
  [arXiv:hep-lat/0312006].
  %%CITATION = PTPSA,153,118;%%

\bibitem{taylor}
   B.~J.~Schaefer, these proceedings;
   F.~Karsch, B.~J.~Schaefer, M.~Wagner, J.~Wambach, in preparation.

%\cite{Schaefer:2007pw}
\bibitem{Schaefer:2007pw}
  B.~J.~Schaefer, J.~M.~Pawlowski and J.~Wambach,
  %``The Phase Structure of the Polyakov--Quark-Meson Model,''
  Phys.\ Rev.\  D {\bf 76}, 074023 (2007)
  [arXiv:0704.3234 [hep-ph]].
  %%CITATION = PHRVA,D76,074023;%%

%\cite{Rossner:2007ik}
\bibitem{Rossner:2007ik}
  S.~Roessner, T.~Hell, C.~Ratti and W.~Weise,
  %``The chiral and deconfinement crossover transitions: PNJL model beyond mean
  %field,''
  Nucl.\ Phys.\  A {\bf 814} (2008) 118
  [arXiv:0712.3152 [hep-ph]].
  %%CITATION = NUPHA,A814,118;%%

%\cite{Fukushima:2008is}
\bibitem{Fukushima:2008is}
  K.~Fukushima,
  %``Critical surface in hot and dense QCD with the vector interaction,''
  Phys.\ Rev.\  D {\bf 78} (2008) 114019
  [arXiv:0809.3080 [hep-ph]].
  %%CITATION = PHRVA,D78,114019;%%

%\cite{Stephanov:2007fk}
\bibitem{Stephanov:2007fk}
  M.~A.~Stephanov,
  %``QCD phase diagram: An overview,''
  PoS {\bf LAT2006} (2006) 024
  [arXiv:hep-lat/0701002].
  %%CITATION = POSCI,LAT2006,024;%%
  
%\cite{Berges:2000ew}
\bibitem{Berges:2000ew}
  J.~Berges, N.~Tetradis and C.~Wetterich,
  %``Non-perturbative renormalization flow in quantum field theory and
  %statistical physics,''
  Phys.\ Rept.\  {\bf 363}, 223 (2002)
  [arXiv:hep-ph/0005122];
  %%CITATION = PRPLC,363,223;%%
%\cite{Gies:2006wv}
%\bibitem{Gies:2006wv}
  H.~Gies,
  %``Introduction to the functional RG and applications to gauge theories,''
  arXiv:hep-ph/0611146.
  %%CITATION = HEP-PH/0611146;%%

%\cite{Roberts:2000aa}
\bibitem{Roberts:2000aa}
  C.~D.~Roberts and S.~M.~Schmidt,
  %``Dyson-Schwinger equations: Density, temperature and continuum strong
  %QCD,''
  Prog.\ Part.\ Nucl.\ Phys.\  {\bf 45}, S1 (2000),
  [arXiv:nucl-th/0005064].
  %%CITATION = PPNPD,45,S1;%%

%\cite{Fischer:2006ub}
\bibitem{Fischer:2006ub}
  C.~S.~Fischer,
  %``Infrared properties of QCD from Dyson-Schwinger equations,''
  J.\ Phys.\ G {\bf 32}, R253 (2006)
  [arXiv:hep-ph/0605173].
  %%CITATION = JPHGB,G32,R253;%%

%\cite{Bender:1996bm}
\bibitem{Bender:1996bm}
  A.~Bender, D.~Blaschke, Y.~Kalinovsky and C.~D.~Roberts,
  %``Continuum study of deconfinement at finite temperature,''
  Phys.\ Rev.\ Lett.\  {\bf 77} (1996) 3724,
  [arXiv:nucl-th/9606006].
  %%CITATION = PRLTA,77,3724;%% 

%\cite{Braun:2006jd}
\bibitem{Braun:2006jd}
  J.~Braun and H.~Gies,
  %``Chiral phase boundary of QCD at finite temperature,''
  JHEP {\bf 0606}, 024 (2006)
  [arXiv:hep-ph/0602226].
  %%CITATION = JHEPA,0606,024;%%

%\cite{Braun:2007bx}
\bibitem{Braun:2007bx}
  J.~Braun, H.~Gies and J.~M.~Pawlowski,
  %``Quark Confinement from Color Confinement,''
  arXiv:0708.2413 [hep-th];
  %%CITATION = ARXIV:0708.2413;%%
%\cite{Marhauser:2008fz}
%\bibitem{Marhauser:2008fz}
  F.~Marhauser and J.~M.~Pawlowski,
  %``Confinement in Polyakov Gauge,''
  arXiv:0812.1144 [hep-ph].
  %%CITATION = ARXIV:0812.1144;%%

%\cite{Braun:2008pi}
\bibitem{Braun:2008pi}
  J.~Braun,
  %``The QCD Phase Boundary from Quark-Gluon Dynamics,''
  arXiv:0810.1727 [hep-ph].
  %%CITATION = ARXIV:0810.1727;%%

%\cite{Fischer:2009wc}
\bibitem{Fischer:2009wc}
  C.~S.~Fischer,
  %``Deconfinement phase transition and the quark condensate,''
  Phys. Rev. Lett. {\bf 103}, 052003 (2009), arXiv:0904.2700 [hep-ph].
  %%CITATION = ARXIV:0904.2700;%%
  
\bibitem{fimu}
 Christian~S.~Fischer and Jens~A.~Mueller, arXiv:0908.0007 [hep-ph].
 %%CITATION = ARXIV:0908.0007;%%

\bibitem{heid}
 J.~Braun, L.~Haas, F.~Marhauser and J.~M.~Pawlowski, 
 arXiv:0908.0008 [hep-ph].
 %%CITATION = ARXIV:0908.0008;%%

%\cite{Gattringer:2006ci}
\bibitem{Gattringer:2006ci}
  C.~Gattringer,
  %``Linking confinement to spectral properties of the Dirac operator,''
  Phys.\ Rev.\ Lett.\  {\bf 97} (2006) 032003,
  [arXiv:hep-lat/0605018].
  %%CITATION = PRLTA,97,032003;%%

%\cite{Bilgici:2008qy}
\bibitem{Bilgici:2008qy}
  E.~Bilgici, F.~Bruckmann, C.~Gattringer and C.~Hagen,
  %``Dual quark condensate and dressed Polyakov loops,''
  Phys.\ Rev.\  D {\bf 77} (2008) 094007,
  [arXiv:0801.4051].
  %%CITATION = PHRVA,D77,094007;%%      

%\cite{Synatschke:2008yt}
\bibitem{Synatschke:2008yt}
  F.~Synatschke, A.~Wipf and K.~Langfeld,
  %``Relation between chiral symmetry breaking and confinement in YM-theories,''
  Phys.\ Rev.\  D {\bf 77} (2008) 114018,
  [arXiv:0803.0271 [hep-lat]].
  %%CITATION = PHRVA,D77,114018;%%

%\cite{Cucchieri:2007ta}
\bibitem{Cucchieri:2007ta}
  A.~Cucchieri, A.~Maas and T.~Mendes,
  %``Infrared properties of propagators in Landau-gauge pure Yang-Mills theory
  %at finite temperature,''
  Phys.\ Rev.\  D {\bf 75}, 076003 (2007),
  [arXiv:hep-lat/0702022].
  %%CITATION = PHRVA,D75,076003;%%

%\cite{Fischer:2008wy}
\bibitem{Fischer:2008wy}
  C.~S.~Fischer and R.~Williams,
  %``Beyond the rainbow: effects from pion back-coupling,''
  Phys.\ Rev.\  D {\bf 78} (2008) 074006,
  [arXiv:0808.3372 [hep-ph]].
  %%CITATION = PHRVA,D78,074006;%%

%\cite{Maris:2000ig}
\bibitem{Maris:2000ig}
  P.~Maris, C.~D.~Roberts, S.~M.~Schmidt and P.~C.~Tandy,
  %``T-dependence of pseudoscalar and scalar correlations,''
  Phys.\ Rev.\  C {\bf 63}, 025202 (2001)
  [arXiv:nucl-th/0001064].
  %%CITATION = PHRVA,C63,025202;%%

%\cite{Horvatic:2007qs}
\bibitem{Horvatic:2007qs}
  D.~Horvatic, D.~Klabucar and A.~E.~Radzhabov,
  %``$\eta$ and $\eta'$ mesons in the Dyson-Schwinger approach at finite
  %temperature,''
  Phys.\ Rev.\  D {\bf 76} (2007) 096009
  [arXiv:0708.1260 [hep-ph]].
  %%CITATION = PHRVA,D76,096009;%%

%\cite{Karsch:1998ua}
\bibitem{Karsch:1998ua}
  F.~Karsch,
  %``Deconfinement and chiral symmetry restoration,''
  arXiv:hep-lat/9903031.
  %%CITATION = HEP-LAT/9903031;%%

%\cite{Roberge:1986mm}
\bibitem{Roberge:1986mm}
  A.~Roberge and N.~Weiss,
  %``GAUGE THEORIES WITH IMAGINARY CHEMICAL POTENTIAL AND THE PHASES OF QCD,''
  Nucl.\ Phys.\  B {\bf 275} (1986) 734.
  %%CITATION = NUPHA,B275,734;%%

%\cite{Fingberg:1992ju}
\bibitem{Fingberg:1992ju}
  J.~Fingberg, U.~M.~Heller and F.~Karsch,
  %``Scaling And Asymptotic Scaling In The SU(2) Gauge Theory,''
  Nucl.\ Phys.\  B {\bf 392} (1993) 493; 
  [arXiv:hep-lat/9208012].
  %%CITATION = NUPHA,B392,493;%%
%\cite{Kaczmarek:2002mc}
%\bibitem{Kaczmarek:2002mc}
  O.~Kaczmarek, F.~Karsch, P.~Petreczky and F.~Zantow,
  %``Heavy quark anti-quark free energy and the renormalized Polyakov loop,''
  Phys.\ Lett.\  B {\bf 543} (2002) 41; 
  [arXiv:hep-lat/0207002].
  %%CITATION = PHLTA,B543,41;%%
%\cite{Lucini:2005vg}
%\bibitem{Lucini:2005vg}
  B.~Lucini, M.~Teper and U.~Wenger,
  %``Properties of the deconfining phase transition in SU(N) gauge theories,''
  JHEP {\bf 0502} (2005) 033; 
  [arXiv:hep-lat/0502003].
  %%CITATION = JHEPA,0502,033;%%

%\cite{Kashiwa:2009ki}
\bibitem{Kashiwa:2009ki}
  K.~Kashiwa, H.~Kouno and M.~Yahiro,
  %``Dual quark condensate in the Polyakov-loop extended NJL model,''
  arXiv:0908.1213 [hep-ph].
  %%CITATION = ARXIV:0908.1213;%%

\end{thebibliography}
\end{document}